\documentclass{aa}  
\usepackage{graphicx}
\usepackage{subfigure}
\usepackage{natbib}
\usepackage{txfonts}
\begin{document} 
   \title{The radio core structure of the luminous infrared galaxy NGC\,4418}
   \subtitle{A young clustered starburst revealed?}
   \author{E. Varenius \inst{\ref{inst:chalmers}}
          \and
          J. E. Conway \inst{\ref{inst:chalmers}}
          \and
          I. Martí-Vidal \inst{\ref{inst:chalmers}}
          \and
          S. Aalto \inst{\ref{inst:chalmers}}
          \and
          R. Beswick \inst{\ref{inst:rob}}
          \and
          F. Costagliola \inst{\ref{inst:granada}}
          \and
          H.-R. Kl{\"o}ckner \inst{\ref{inst:hrk}}
          }
   \institute{
              Department of Earth and Space Sciences,
              Chalmers University of Technology, 
              Onsala Space Observatory,
              349 92 Onsala, 
              Sweden \\
              \email{varenius@chalmers.se}
              \label{inst:chalmers}
              \and
              Jodrell Bank Centre for Astrophysics,
              Alan Turing Building, 
              School of Physics and Astronomy, \\
              The University of Manchester,
              Manchester M13 9PL, 
              UK
              \label{inst:rob}
              \and
              Instituto de Astrofísica de Andalucía, 
              Glorieta de la Astronomía, 
              s/n, 
              18008 Granada, 
              Spain
              \label{inst:granada}
              \and
              Max-Planck-Institut f\"ur Radioastronomie, Auf dem H\"ugel 69, 53121 Bonn, Germany
              \label{inst:hrk}
              }
   \date{Received 20 December 2013 / Accepted 13 February 2014}
  \abstract
      {
       The galaxy NGC\,4418 contains one of the most compact obscured nuclei
       within a luminous infrared galaxy (LIRG) in the nearby Universe. This
       nucleus contains a rich molecular gas environment and an unusually high
       ratio of infrared to radio luminosity (q-factor). The compact nucleus is
       powered by either a compact starburst or an active galactic nucleus
       (AGN). 
      }
      {
       The aim of this study is to constrain the nature of the nuclear region
       (starburst or AGN) within NGC\,4418 via very-high-resolution radio imaging.
      }
      {
       Archival data from radio observations using the European Very Long
       Baseline Interferometry Network (EVN) and Multi-Element Radio Linked
       Interferometer Network (MERLIN) interferometers are imaged.  Sizes and flux
       densities are obtained by fitting Gaussian intensity distributions to the image.
       The average spectral index of the compact radio emission is estimated from 
       measurements at 1.4\,GHz and 5.0\,GHz.
      }
      {  
       The nuclear structure of NGC\,4418 visible with EVN and MERLIN consists
       of eight compact (<49\,mas i.e. <8\,pc) features spread within a region
       of 250\,mas, i.e. 41\,pc.  We derive an inverted spectral index
       $\alpha\ge0.7$ ($S_\nu\propto\nu^{\alpha}$) for the compact radio
       emission.
      }
   {
       Brightness temperatures $>10^{4.8}$\,K indicate that these compact
       features cannot be HII-regions.  The complex morphology and inverted
       spectrum of the eight detected compact features is evidence against the
       hypothesis that an AGN alone is powering the nucleus of NGC\,4418. The
       compact features could be super star clusters (SSCs) with intense star
       formation, and their associated free-free absorption could then
       naturally explain both their inverted radio spectrum and the low radio
       to IR ratio of the nucleus. The required star formation area density is
       extreme, however, and close to the limit of what can be observed in a
       well-mixed thermal/non-thermal plasma produced by star-formation, and is
       also close to the limit of what can be physically sustained. 
   }

   \maketitle
%

\section{Introduction}
The galaxy \object{NGC\,4418} (IRAS12243-0036) is a luminous
($L_{FIR}>10^{11}L_\sun$) infrared galaxy (LIRG) 
with an infrared flux density more than 5 times larger than expected from the
linear radio to far-infrared relation.  \cite{yun} find only ten such objects
in a sample of 1809 galaxies, which makes NGC\,4418 a very unusual object.
The galaxy has one of the deepest mid-IR silicate absorption features ever
detected towards an external galaxy indicating a very deeply obscured nucleus.
The deep absorption and high IR brightness of NGC\,4418 was first noted by
\cite{roche1986} who proposed that this galaxy hides either an active
galactic nucleus (AGN) or a compact nuclear starburst.
Very faint H$\alpha$ emission has been detected, but the absence of NII,
OI, and SII emission makes it difficult to classify the galaxy as an AGN or
starburst based on its optical spectrum \citep{armus1989}.  Furthermore, near-IR
and mid-IR observations are consistent with both hypotheses \citep{evans2003}.

Millimetre and submm observations by \cite{sakamoto2013} and
\cite{francescolatest} reveal the presence of a highly compact (<0."1) high
surface brightness continuum source suggesting that the bulk of the galaxy FIR
emission emerges from a source less than 20\,pc in diameter.  The extreme
inferred H$_2$ column density $>10^{25}\text{cm}^{-2}$ towards this nucleus 
\citep{gonzales2012} makes it extremely difficult to determine the nature of the
buried source.

Despite extensive studies of NGC\,4418, the nature of its central power source
is not clear.  High-resolution radio very long baseline
interferometry (VLBI) observations provide a possible way to distinguish
between AGNs and starbursts.  If the source of emission is an AGN this may
produce high brightness temperature compact radio components.  If the source of
emission is a young starburst this would instead produce multiple supernovae
(SNe) and supernova remnants (SNRs).  A mix between these two scenarios, i.e.
AGN and SNe/SNRs, is also possible.

Radio VLBI observations with (sub)milli arcsecond resolution have proven to be a
valuable tool to distinguish between these scenarios by directly probing 
the central regions (U)LIRGs, e.g. for Arp\,220 \citep{batejat2011} and Arp\,299
\citep{Arp299Paper}.
In this paper we report on the analysis of archival NGC\,4418 data from the EVN and
MERLIN interferometers which for the first time reveal eight discrete
compact radio features within its nucleus.

In Sect.\,\ref{sec:cal} we summarise the data used and the calibration
procedures applied.  In Sect.\,\ref{sec:imaging} we describe the imaging
process and discuss the image fidelity.  In Sect.\,\ref{sec:results} we present
the results of the imaging and simple modelling.  In
Sect.\,\ref{sec:discussion} we briefly discuss the hypotheses of AGN/starburst
in relation to our results.  Finally, in Sect.\,\ref{sec:conclusions} we
summarise our conclusions. In this paper we assume a distance of 34\,Mpc to
NGC\,4418 at which an angular size of 1 mas corresponds to 0.165\,pc
\citep{francescolatest}.


\section{Data and calibration}
\label{sec:cal}
We analysed archival data from a combined EVN and MERLIN experiment (EB019, P.I.:
P. Barthel) observed at frequency 5.0\,GHz taken on June 3, 2001. The observation
spanned 5.5 hours with 2.8 hours integration time on NGC\,4418.
These data were analysed in two ways: using only the MERLIN data, and 
as a combined dataset with both EVN and MERLIN visibilities.
Below in Sects. \ref{sec:calmerlin} and \ref{sec:calevn} we describe the calibration 
of the EVN and MERLIN data. In Sec. \ref{sec:combine} we then present the details of
how images were obtained from the MERLIN-only and EVN and MERLIN data sets.

\subsection{Calibration of the MERLIN data}
The MERLIN observations were taken without the Lovell Telescope. The 
data were calibrated using standard procedures. The sources J1232-0224 and J1229+0203
were used as phase calibrators, 3C\,286 as primary flux density calibrator,
J0555+3948 as secondary flux density calibrator, and OQ208 to derive bandpass
solutions. These data have been published before 
by \cite{francescolatest}. We use the same calibrated dataset here and for further 
details of the calibration process we refer the reader to Sect.\,2.1 of \cite{francescolatest}.
\label{sec:calmerlin}

\subsection{Calibration of the EVN data}
\label{sec:calevn}
The EVN observations included eight antennas (only a part of the full EVN):
Effelsberg (100m diameter), Jodrell Bank Mk2 (25m), Medicina (32m), Noto
(32m), Torun (32m), Westerbork (14x25m), Onsala (25m), and Cambridge (32m). 

The source J1229+0203 (3C273), 3.0$^\circ$ from NGC\,4418, was observed once every 25
minutes to track instrumental delay and rate variations and J1232-0224,
2.0$^\circ$ from NGC\,4418, was observed as phase calibrator in duty cycles of
4min/2.5min with NGC\,4418.  The parallel-hand products (RR and LL) were
correlated and recorded in four sub-bands (IFs) of 16 channels each containing a
total bandwidth of 32\,MHz.

Calibration of the EVN dataset was done using standard procedures within the
Astronomical Image Processing System (AIPS) \citep{greisen}, ParselTongue 2.0
\citep{Kettenis}, and Difmap \citep{shepherd}.  First, bandpass corrections were
derived using J1229+0203.  Then, bad data were edited using standard procedures
in AIPS.  After editing, residual delays and rates were corrected for on
J1229+0203 and J1232-0224, using the global fringe fitting algorithm (see e.g.
\citealt{thompson}) as implemented in AIPS.  The flux scale was set from a
priori gain and system temperature measurements.  

Since J1232-0224 is resolved at EVN baselines, the data were loaded into Difmap
and phase solutions found by executing several self-calibration/clean
iterations with an averaging time of 10\,min, followed by amplitude and phase
self-calibration with an averaging time of 30\,min.  Then, amplitude self-calibration 
was done on J1232-0224 in AIPS using averaging times of 100\,minutes
to correct for obvious global offsets in the a priori amplitude calibration of
a subset of antennas. Finally, the amplitude and phase corrections derived on the
calibrator were transferred and applied to the target source. No self-calibration 
was done using the target source itself.

\subsection{Combining the data from EVN and MERLIN}
\label{sec:combine}
Once the 5\,GHz EVN and MERLIN observations had been independently calibrated,
the data were exported to the program CASA \citep{CASA}, version 3.4, and
concatenated into one measurement set using the task CONCAT.  We then used the
ms-tool in CASA to scale the amplitudes of the EVN visibilities to those of
MERLIN at similar baseline lengths as done previously by \cite{perucho2012}.
To ensure positional alignment the phase calibrator J1232-0224 was checked to
have the same coordinates in both the EVN and the MERLIN data sets; after
imaging J1232-0224 the peak was found to be at R.A. 12h32m0.016s, Dec.
-02$^\circ$24'04''.770 in both data sets. 

\section{Imaging}
\label{sec:imaging}
In this section we describe the deconvolution of the MERLIN-only data and of
the combined EVN and MERLIN data. We also discuss the fidelity of the
EVN and MERLIN image.

\subsection{MERLIN-only at 5\,GHz}
\begin{figure}[htbp]
\centering
    \includegraphics[width=0.5\textwidth]{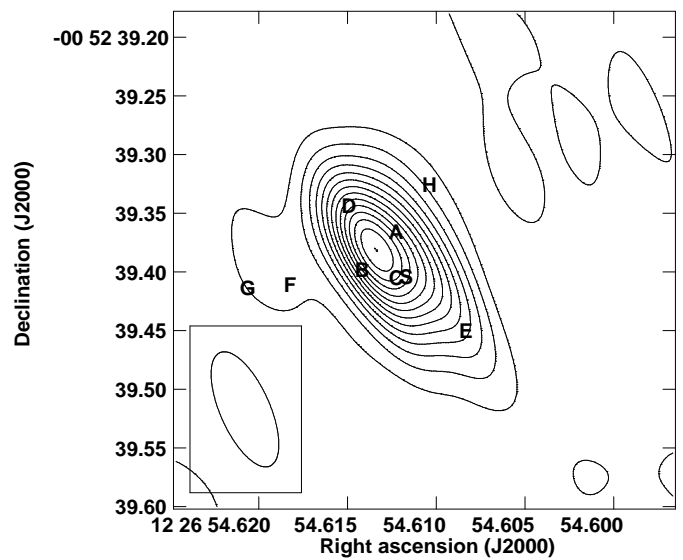}
\caption{
  MERLIN-only continuum image of NGC\,4418 at 5.0\,GHz. Contour levels are 
    (-2,2,4,6,8,10,12,14,16,18,20,22,24,26,28) times
    the noise RMS $\sigma=300$\,$\mu$Jy/beam.
    We note that there are no negative features as strong as -2$\sigma$, hence no
    dashed contours.  The CLEAN restoring PSF of 
    FWHM 105\,mas$\times$42\,mas is plotted in the lower left.
    The labels A-H mark the positions of the compact features seen in Fig.
    \ref{fig:wt1.0}.
    The label S (close to label C) corresponds to the
    860\,$\mu$m continuum peak position from \citet{sakamoto2013}.  
}
    \label{fig:MERLIN5}
\end{figure}

An image at 5.0\,GHz using data from MERLIN-only (no EVN)  was obtained using
the CLEAN algorithm as implemented in CASA 3.4 and is shown in Fig.
\ref{fig:MERLIN5}.  The calibrated visibilities were weighted relative to each
other using the Briggs weighting scheme
\citep{briggs} with robust parameter 0.  

\subsection{EVN and MERLIN at 5\,GHz}
\begin{figure}[htbp]
\centering
    \includegraphics[trim=0cm 0cm 0cm 0cm, clip=true, width=0.5\textwidth]{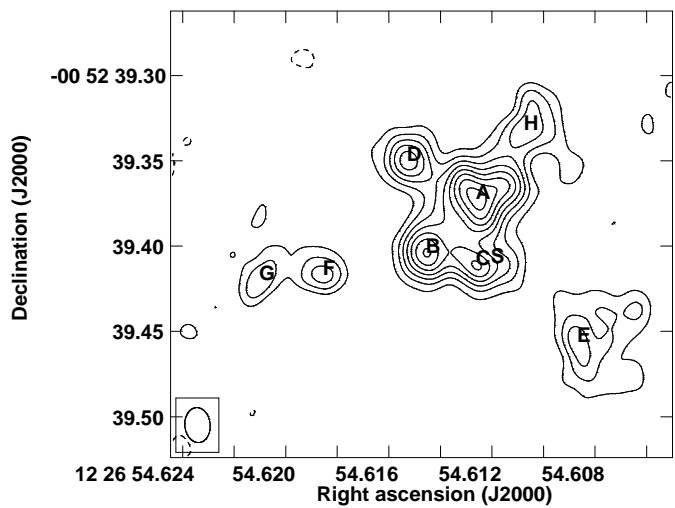}
\caption{
    EVN and MERLIN 5\,GHz image of NGC\,4418 with relative weight 1.0 between the
    EVN and MERLIN data. The contour levels are (-2,2,4,6,8,10,12,14,16)
    times the RMS noise of $\sigma=90\mu$\,Jy/beam.  Eight compact features are
    visible above 5$\sigma$ labelled from A to H. 
    The label S (close to label C) corresponds to the
    860\,$\mu$m continuum peak position from \citet{sakamoto2013}.  
    The CLEAN
    restoring PSF of FWHM 20.6\,mas$\times$14.8\,mas is plotted in the lower left.
}
\label{fig:wt1.0}
\end{figure}

The combined dataset was imaged using the deconvolution algorithm CLEAN as
implemented in CASA 3.4. To maximise the sensitivity the
calibrated visibilities were tapered using a 7.5M$\lambda$ Gaussian taper and
weighted using natural weighting. 
The multi-scale option was used in CLEAN to improve deconvolution of more
extended features. Careful conservative masking was done to minimise the risk
of creating spurious sources from side lobes in the PSF.
Figure \ref{fig:wt1.0} shows the deconvolved image.

\begin{figure}[htbp]
\centering
    \includegraphics[width=0.4\textwidth]{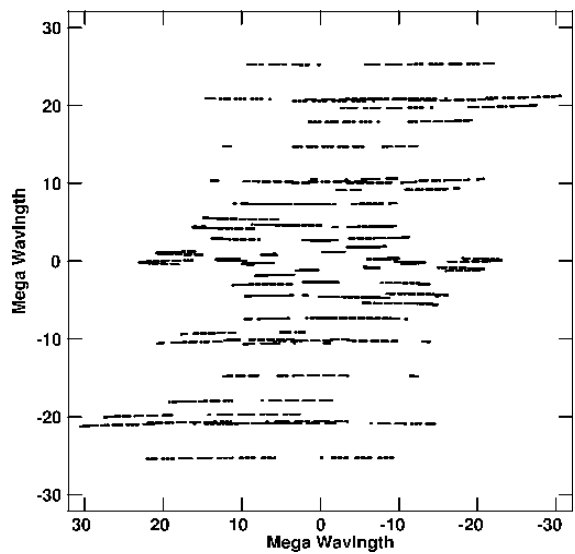}
\caption{
    The coverage of the Fourier plane for the EVN data, plotted as V vs U in
    units of M$\lambda$. The discrete set of spatial scales sampled in
    delincation is mainly due to the small (close to 0) declination of
    NGC\,4418.  The concatenation of the MERLIN data adds shorter baselines in
    the centre of less than about 3\,M$\lambda$.
    \label{fig:UVEVN}
}
\end{figure}

Because of the small (close to 0) declination of NGC\,4418, limited bandwidth,
and missing EVN antennas the observations sample a discrete set of spatial
scales in declination (see Fig.  \ref{fig:UVEVN}). This results in a point
spread function (PSF) with strong side lobes in declination, reaching in
amplitude up to 30\% of the peak.  When the PSF has strong side lobes there is
risk of creating spurious sources in the deconvolution process.  To assess the
image quality, we performed tests by changing the relative weighting of the
data before imaging.  

\begin{figure}[htbp]
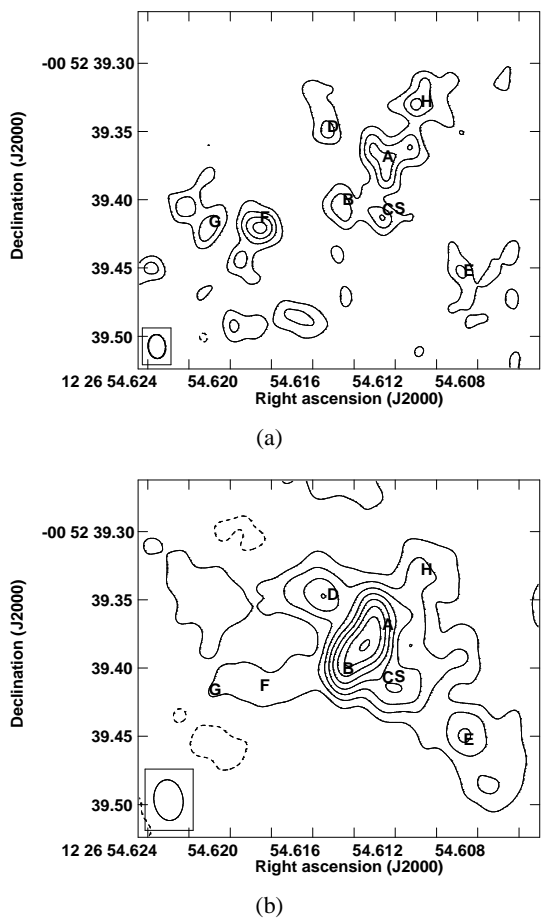

\centering
\subfigure[]{
    \includegraphics[width=0.4\textwidth]{figs/WT01-RMS90.EPS}
    \label{fig:wt0.1}
}
\subfigure[]{
    \includegraphics[width=0.4\textwidth]{figs/WT10-RMS90.EPS}
    \label{fig:wt10}
}
\caption[Optional caption for list of figures]
{
    5\,GHz EVN and MERLIN images of NGC\,4418 with different relative weighting. In
    \subref{fig:wt0.1} the weight of the EVN is increased by a factor of ten
    with respect to MERLIN.
    As expected, the features F and G are still prominent, but the features
    A, B, C, D, E, and H are weaker.  In \subref{fig:wt10} the weights of MERLIN
    data are increased by a factor of ten with respect to the EVN.  
    F and G are now lost in the noise, but we still
    see A, B, C, D, E, and H.  The noise RMS increases to 95\,$\mu$Jy/beam for Fig.
    \subref{fig:wt0.1} and 150\,$\mu$Jy/beam for \subref{fig:wt10}.  For
    easy comparison the contours in Figs.  \subref{fig:wt0.1} and \subref{fig:wt10}
    have been set to the same (-2,2,4,6,8,10,12,14,16) times 90\,$\mu$Jy as in
    Fig. \ref{fig:wt1.0}.
}
\label{fig:not1.0}
\end{figure}

Increasing the relative weight of the EVN should significantly reduce the flux
densities of the resolved features A, B, C, D, E, and H compared to the
unresolved features F and G.  Increasing the relative weights of the EVN data
by a factor of ten with respect to MERLIN we obtained what we expected (see
Fig. \ref{fig:wt0.1}).  
The reverse experiment, increasing the relative weights of MERLIN data to ten
times the EVN, should result in increased flux densities of A, B, C, D, E, 
and H compared to F and G.  Again, we obtained what we expected, see Fig.
\ref{fig:wt10}.  The lowest RMS noise is obtained when using the same weight
for EVN and MERLIN (see Fig.  \ref{fig:wt1.0}). We are confident that the
compact features detected here are real.

\subsection{Estimates of uncertainties}
\label{sec:uncertainties}
We estimate the absolute flux density calibration to be accurate to within 10\%
as given by the MERLIN staff.  We assume the EVN flux density scale to
be as accurate as MERLIN since we aligned the amplitudes at similar baseline
lengths. To account for possible flux density errors in the deconvolution
process (e.g. masking effects) we add another 5\% uncertainty giving a total
accuracy of 15\% for measured flux densities.

We estimate an absolute positional calibration accuracy of <1\,mas.  Given the
EVN and MERLIN beam size and image dynamic range of $\sim 17$ we estimate a
positional uncertainty of 3\,mas due to the image noise.  We conservatively
estimate a total positional uncertainty of 4\,mas (a fifth of the major axis
of the beam) for the EVN-MERLIN image. Similarly, we estimate an uncertainty of
20\,mas for the MERLIN-only peak position.

\section{Results and modelling}
\label{sec:results}
In this section we present the results obtained from the deconvolved images and
from simple Gaussian modelling. All measured and fitted quantities are
summarised in Tables \ref{tab:results} and \ref{tab:props}.

\subsection{Description of the detected structure}
\label{sec:fitting}
Eight compact features are detected brighter than $5\sigma$ and are 
shown in Fig.  \ref{fig:wt1.0}. For reference
each feature is labelled from A to H.  At 5.0\,GHz the synthesised MERLIN-only
PSF is relatively large so most of the compact features blend together (Fig.
\ref{fig:MERLIN5}).  We note, however, that the position of E matches very well
with the tail visible to the south-west in Fig. \ref{fig:MERLIN5}.  Features F
and G are farther away from the rest and, although there are indications, 
are too weak to be seen clearly with MERLIN-only.

The label S in Figs. \ref{fig:MERLIN5} and \ref{fig:wt1.0} marks the
position of the 860\,$\mu$m peak found by \cite{sakamoto2013}. The PSF of
these submillimetre wavelength observations had a size of 690\,mas$\times550$\,mas
which encompasses all eight of the features we see.
Therefore, we cannot say that the 860\,$\mu$m position is associated with one
single radio component. 
We, note, however that the deconvolved size of FWHM 0".10 derived from the
860\,$\mu$m continuum is very similar to the size of the region occupied by the
four brightest radio features labelled A-D in Fig. \ref{fig:wt1.0}.

\begin{table*}[htbp]
    \caption{Data for deconvolvd images of NGC\,4418}
    \label{tab:results}
    \centering
    \begin{tabular}{l c c}
        \hline\hline
         & MERLIN 5\,GHz  & EVN and MERLIN 5\,GHz \\
        \hline
        Point Spread Function FWHM [mas] & 105$\times$42 & $20.6\times14.8$ \\
        Point Spread Function pos. angle [$^\circ$] & 23.6 & 5.2 \\
        Image noise RMS $\sigma$ [mJy/beam] & 0.3 & 0.09\\ 
        Integrated flux density S$_I$ [mJy] & $17.7\pm2.7$ & $22.0\pm3.3$ \\ 
        Peak brightness $P$ [mJy/beam] &  8.40$\pm$1.30 & 1.50$\pm$0.24\\ 
        Peak R.A. (J2000)  & 12h26m54.6133s$\pm0.0013$s& 12h26m54.6126s$\pm0.0003$s\\ 
        Peak Dec. (J2000) & -00$^\circ$52'39''.382$\pm20$\,mas & -00$^\circ$52'39''.372$\pm4$\,mas\\ 
        \hline                                
    \end{tabular}
    \tablefoot{
        The integrated flux densities were obtained by summing
        all pixels above $3\sigma$. The PSF given is the restoring beam as
        calculated and used by the CLEAN algorithm. Uncertainties on peak brightness include
        both calibration and image noise as $\sqrt{(0.15P)^2+\sigma^2}$. 
    }
\end{table*}

\setlength{\tabcolsep}{3pt}
\begin{table*}[htbp]
    \caption{Properties of identified compact features, labelled as in Fig. \ref{fig:wt1.0}.}
    \label{tab:props}
    \centering
    \begin{tabular}{l c c c c c c c c }
        \hline\hline
        Label in Fig. \ref{fig:wt1.0}: & A & B & C & D & E & F & G & H\\
        \hline
        \textbf{Measured peak values} & & &  &  & & & & \\
        R.A. 12h26m54.[s] & 6126$\pm3$ & 6145$\pm3$ & 6126$\pm3$ & 6153$\pm3$ & 6087$\pm3$ & 6185$\pm3$ & 6210$\pm3$ & 6107$\pm3$\\
        Dec. -00$^\circ$52'39''.  & 372$\pm4$ & 404$\pm4$ & 411$\pm4$ & 350$\pm4$ & 456$\pm4$ & 417$\pm4$ & 420$\pm4$ & 332$\pm4$\\
        Brightness [mJy/b.] & 1.50$\pm0.24$ & 1.30$\pm0.21$ & 1.10$\pm0.20$ & 1.00$\pm0.17$ & 0.70$\pm0.14$ & 0.60$\pm0.13$ & 0.50$\pm0.12$ & 0.70$\pm0.14$\\
        \hline
        \textbf{Fitted peak values} & & &  &  & & & & \\
        R.A. 12h26m54.[s] & 6125$\pm3$ & 6147$\pm3$ & 6126$\pm3$ & 6153$\pm3$ & 6080$\pm3$ & 6187$\pm3$ & 6209$\pm3$ & 6105$\pm3$\\
        Dec. -00$^\circ$52'39''. & 371$\pm4$ & 404$\pm4$ & 410$\pm4$ & 349$\pm4$ & 455$\pm4$ & 415$\pm4$ & 417$\pm4$ & 328$\pm4$\\
        Brightness [mJy/b.] & 1.54$\pm0.24$ & 1.15$\pm0.19$& 0.99$\pm0.17$ & 0.97$\pm0.66$ & 0.51$\pm0.11$ & 0.67$\pm0.13$ & 0.55$\pm0.12$ & 0.68$\pm0.13$\\
        \hline
        \textbf{Fitted integrated values} & & &  &  & & & & \\
        S$_I$ [mJy] & 6.99$\pm1.11$ & 2.17$\pm0.36$& 2.98$\pm0.51$ & 2.01$\pm0.34$ & 3.73$\pm0.81$ & 0.954$\pm0.182$ & 0.871$\pm0.182$ & 1.92$\pm0.36$\\
        $\theta_M$ [mas] & 37.8$\pm1.1$& 17.9$\pm1.7$& 33.2$\pm1.8$& 20.2$\pm2.0$& 49.0$\pm2.5$& 0.0 &0.0 &29.3$\pm2.4$\\
        $\theta_m$ [mas] & 27.0$\pm1.0$ & 15.0$\pm2.2$ & 15.0$\pm2.4$ & 14.5$\pm3.0$ & 39.0$\pm3.5$ & 0.0 & 0.0 & 17.8$\pm3.5$\\
        Pos. angle [$^\circ$] & 108$\pm179$ & 160$\pm178$ & 76.5$\pm2.7$ & 81$\pm20$ & 4.5$\pm1.8$ & - & - & 147$\pm178$\\
        \hline
        \textbf{Derived values} & & &  &  & & & & \\
        Size [pc] & 6.2$\times$4.5 & 3.0$\times$2.5 & 5.5$\times$2.5&3.3$\times$2.4&8.1$\times$6.4 & <3.4$\times$2.4& <3.4$\times$2.4& 4.8$\times$2.9\\
        $T_b [\text{log}_{10}[K]]$ & 5.35& 5.42& 5.29& 5.35& 4.8& >5.01& >4.97& 5.08\\
        \hline                                
    \end{tabular}
    \tablefoot{
        \textbf{Positions} are given in the J2000 reference frame.
        Uncertainties in positions due to the Gaussian fitting are small
        compared to the 4\,mas estimated in Sect. \ref{sec:uncertainties}.
        \textbf{Uncertainties } of fitted peak brightness and integrated flux densities include both estimates of 
        uncertainties from calibration/imaging and from the Gaussian fitting as $((0.15S)^2+\sigma_\mathrm{fit}^2)^{0.5}$.
        Uncertainties for fitted sizes and position angles are given as reported by
        IMFIT in CASA 3.4, i.e. without accounting for possible calibration/imaging errors.
        F and G were fitted as point sources so the PSF of
        20.6$\times$14.8\,mas was used as an upper size limit.  
        \textbf{The brightness temperature} at 5.0\,GHz was calculated from the
        fitted integrated flux density $S_I$ assuming a uniform elliptical
        source component with FWHM fitted angular diameters $\theta_M,\theta_m$
        and integrated flux density $S_I$ using Eq. 5 by \cite{condon1991}:
    $T_b=(c^2S_I/2k\nu^2)\cdot(8\text{ln}(2)/3\pi\theta_M \theta_m)$.  
}
\end{table*}

The eight detected compact features were modelled using Gaussian
intensity distributions and fitted simultaneously to the image using the
task IMFIT in CASA 3.4. From IMFIT it was possible to get estimates on the peak
and integrated flux densities as well as estimates for the deconvolved
(intrinsic) sizes for the eight compact features. 
The parameters of the best model are given in Table 
\ref{tab:props}.  The best model recovers 21.6\,mJy which is almost all the
flux density of 22.0\,mJy found in Fig. \ref{fig:wt1.0} by summing all pixels
above $3\sigma$.  After subtracting the model from the image the maximum
residual is 0.4\,mJy/beam, i.e. less than $5\sigma$.  

\subsection{The average spectrum of the compact features}
\label{sec:alpha}
\cite{condon1990} find a total flux density of about 40\,mJy at 1.49\,GHz with
three different resolutions of 18", 5", and 1.5" using the Very Large Array.
From their highest resolution data they estimate a size of 500$\times$300\,mas
for the nucleus of NGC\,4418.
\cite{francescolatest} find 38\,mJy at 1.41\,GHz with higher resolution of
350$\times$160\,mas using MERLIN. 
Imaging and visibility fitting clearly shows a smooth extended component of FWHM $500$\,mas,
all in good agreement with the previous VLA results.
The similar total flux density recovered using MERLIN and the three resolutions of the VLA
strongly indicates that there is no evidence for 1.4\,GHz emission on scales larger than
$\sim$500\,mas.
However, from the amplitude vs UV-distance plot by \cite{francescolatest}
(their Fig A.1) it is clear that the total flux density of the 500\,mas
component detected at 1.4\,GHz is $\sim$29\,mJy. Subtracting this from the
total flux density of 38\,mJy gives 9\,mJy in compact features, unresolved with
MERLIN at 1.4\,GHz. 

Our modelling (see Sect. \ref{sec:fitting}) shows that all 22\,mJy detected in
Fig. \ref{fig:wt1.0} comes from the eight compact features in Fig.
\ref{fig:wt1.0}, each with sizes significantly smaller than the extended 500\,mas
component detected at 1.4\,GHz. We conclude that the emission we see 
is not associated with the extended emission, and that we should compare
the 22\,mJy in compact emission at 5.0\,GHz to the 9\,mJy found at 1.41\,GHz. 
Modelling the spectrum as a simple power law ($S_\nu\propto\nu^{\alpha}$)
we derive an inverted spectral index of $\alpha\ge0.7$ for the compact emission.
This inverted spectrum is very unusual.
Unfortunately we lack high-resolution images at other radio frequency bands
than 5\,GHz and therefore we cannot investigate the radio spectra of
each compact feature in Fig. \ref{fig:wt1.0}. 

\section{Discussion}
\label{sec:discussion}
The object NGC\,4418 is a luminous infrared galaxy with 
very unusual properties such as very weak radio compared to IR,
very high gas density, and very high dust obscuration.  \cite{gonzales2012} proposed
a layered-temperature model for the nucleus of NGC\,4418. Three layers of
different temperature on scales of 100, 20, 
and 5\,pc is also consistent with the MERLIN and SMA observations presented by
\cite{francescolatest}.  
Based on SMA data \cite{sakamoto2013} present strong evidence for a compact
($\sim$20\,pc) core inside a 100\,pc scale concentration of molecular gas.
\cite{francescolatest}
based on 1\,mm SMA data find an even smaller (<5\,pc) hot
($\sim300$K) component in the centre, traced by rotational and vibrational
excitation of the molecules HC$_3$N and HNC, in good agreement with the <5\,pc
component inferred from Hershel mid-IR observations presented by 
\cite{gonzales2012}.  To explain the compact emission it has been suggested
that NGC\,4418 hosts either a radio-weak active galactic nucleus
(AGN) or a young compact starburst (see e.g.  \cite{spoon2001} and
\cite{roussel2003}). A combination of both is also possible. 

In this paper we resolve for the first time the structure in the very centre
of NGC4418 into discrete objects.  We detect eight compact features within a
nuclear region of size 42\,pc (see Fig.  \ref{fig:wt1.0} and Table
\ref{tab:props}). 
For the brightest feature A (7\,mJy) we find a deconvolved area 28.8\,pc$^2$ in good 
agreement with the hot 25\,pc$^2$ component proposed by
\cite{francescolatest} and \cite{gonzales2012}.
Next, the four brightest features (A-D) together contain more than half of the
total flux density (14 of 22\,mJy) detected at 5.0\,GHz in an area (using Eq. 4 by
\cite{condon1991} and the sizes in Table \ref{tab:props}) of about 100\,pc$^2$.
This roughly fits the area of $\pi10^2=310$pc$^2$ suggested by
\cite{sakamoto2013}. 

As noted in Sect.\,\ref{sec:alpha} an extended 500\,mas scale radio component is
found at 1.41\,GHz by \cite{francescolatest} corresponding to a linear size of
80\,pc. This larger scale structure together with the four weakest features
E, F, G, and H (spread over a region of $\sim$40\,pc) in Fig. \ref{fig:wt1.0}
could very well be the structure responsible for the outer layer of size
$\sim100$\,pc traced by CO and HI emission \citep{francescolatest}.
The 22\,mJy we find at 5.0\,GHz can also be compared with the the 26.1\,mJy 
obtained previously at 4.8\,GHz with the VLA by \cite{baan2006}. Since the VLA
is sensitive to much larger spatial scales, this could indicate the presence of an 
extended component at 5.0\,GHz, possibly the same 500\,mas structure detected
at 1.4\,GHz.

In the following sections we briefly discuss the observed multiple radio
features and the overall inverted spectrum in relation to the AGN/starburst models
for the power source of NGC\,4418.

\subsection{An active galactic nucleus}
Several authors have suggested that an AGN is present in the nucleus of
NGC\,4418, mainly because of the high IR brightness and lack of significant
emission from polycyclic aromatic hydrocarbon (PAH), see e.g.
\cite{spoon2001}.
\cite{maiolino2003} observed NGC\,4418 with Chandra finding "evidence for a
flat spectrum emission component which may imply the presence of a
Compton-thick AGN, but the limited photon statistics make this identification
somewhat tentative". A Compton thick AGN would be consistent with the 
high H$_2$ column density of 
$>10^{25}\text{cm}^{-2}$ inferred by \cite{gonzales2012}.
If the nucleus of NGC\,4418 does harbour a radio-emitting AGN we expect a 
linear radio structure, for example jets or lobes straddling a central accretion region.  
Instead, we see a rich non-linear structure of several features in Fig. \ref{fig:wt1.0}, 
which argues against the radio emission coming from \emph{only} an AGN. 
We note that the complex structure is \emph{not evidence against} an AGN;
some part of the structure (e.g. the linear structure B-A-H) could be due to
an AGN, but an AGN alone cannot explain \emph{all} the features we see.

The inverted radio spectrum could be caused by attenuation of synchrotron
radiation from an AGN at 1.4GHz due to either synchrotron self-absorption or
foreground thermal (free-free) absorption.  At GHz frequencies synchrotron
self-absorption is only important for sources with brightness temperatures
$>10^{10}$K (e.g.  \cite{condon1992} Sect. 4.1), i.e. much higher than the
$\sim10^5$ derived in Table \ref{tab:props} for the bright resolved features,
for example A-D.  There are cases of thermal absorption in AGN where the
absorption is suggested as due to the subparsec scale accretion disk, for
example NGC\,1068 \citep{gallimore} and NGC\,4261 \cite{jones2001}. This
scenario is an unlikely explanation for the overall spectrum of the nucleus of
NGC\,4418, where the compact radio emission is distributed over tens of
parsecs.

\cite{gallimore2004} present radio VLBI observations at several epochs and
frequencies of the LIRG NGC\,6240 which appears to host a pair of AGNs
 in addition to a starburst. The radio-morphology of NGC\,6240 is different
from NGC\,4418, with one source being compact ($<$2\,pc) and the other showing a
linear structure common for AGN. Both AGN have inverted spectra below 2.4\,GHz,
suggesting strong free-free absorption on scales smaller than 25\,pc. It is not
clear if this relatively large absorbing medium is associated with the two AGNs
or if it is due to star formation between the two nuclei.

To our knowledge there is no clear evidence of thermal absorption in AGNs on
scales of tens of parsec required here without a coexistent burst of star
formation.  This implies that the observed inverted radio spectrum seen in the
nucleus of NGC\,4418, like the radio morphology, cannot be \emph{only due to}
an AGN and implies that there must be a strong starburst component.

\subsection{Starburst}
A starburst will produce HII-regions, supernovae (SNe) and supernova
remnants (SNRs). 
The HII-regions can be radio bright due to thermal (free-free) emission from
the ionised gas, with brightness temperatures up to $2\cdot10^4$\,K (see e.g.
Sect.\,11.2.1 in \cite{tools}).
Most of the compact features detected in Fig. \ref{fig:wt1.0} have brightness
temperatures above $10^{5}$\,K (only E and G are below $10^{5}$\,K, see
Table \ref{tab:props}), effectively ruling out the possibility of the features
A-H being HII-regions.

\subsubsection{Single SNe/SNRs}
Supernovae and supernova remnants can be radio bright because of synchrotron
emission in the expanding shell.
The environment in the nucleus of NGC\,4418 is thought to be similar (e.g. the
high density) to what is found in the nuclear regions of Arp\,220.
If the features A-H were single SNRs we would expect them to follow the
luminosity-size relation as SNe/SNRs do in Arp\,220 and M82 
\citep[see Fig. 5 in][]{batejat2011}. 
Assuming a typical spectral index observed for an SNR in Arp\,220 \cite{batejat2011}, 
this implies an equivalent 3.6\,cm spectral luminosity of
$L_{3.6\text{cm}}=1700\cdot10^{24}$ergs/s/Hz for a typical 2\,mJy (at 6\,cm)
feature in NGC\,4418 at distance 34\,Mpc.  This corresponds to SNRs in Arp\,220
of size $\approx0.3$\,pc, i.e. significantly smaller than sizes observed for
A, B, C, D, E, and H (listed in Table \ref{tab:props}), strongly indicating that
these features are not single SNe/SNRs.  Features F and G are unresolved in 
Fig. \ref{fig:wt1.0} and are still compatible with being SNe/SNRs. 

\subsection{Super star clusters}
While the brightest features are probably not \emph{single} SNe/SNRs they might be
super star clusters (SSCs) with intense star formation.  \cite{sakamoto2013}
note that a starburst in nucleus of NGC\,4418 could be in the form of massive
star clusters compatible with the observed high luminosity/mass ratio of
500-1000$L_\sun M_\sun^{-1}$. This high L/M ratio is comparable to the
theoretical maximum of 500-1000$L_\sun M_\sun^{-1}$ for a cluster-forming gas
cloud or disk \citep[see][]{scoville2003,thompson2005}.  This limit arises
because too much radiation pressure will blow away the star forming gas and
halt the star formation. The limit is, however, strictly valid for steady state
star formation, and it is possible to get even higher ratios for a short while.
The observed ratio for NGC\,4418 is high, but does not rule out star formation
powering the nucleus.

The size of an SSC can be estimated from its mass using the equation
$r_{cl}\sim0.3(M_\text{cl}/10^6M_\sun)^{3/5}\text{pc}$ by \cite{murray2009}.
Assuming the total SSC mass of $M_\text{dyn}-M_\text{bh}-M_\text{mol}=9\cdot10^7M_\sun$ 
(with values from Table 8 by \cite{sakamoto2013}) is shared by the eight detected features
according to their relative 5\,GHz luminosity, we estimate
SSC diameters of 1.3-4.5\,pc, in good agreement with the observed sizes.

Young massive SSCs of similar mass have been detected by optical imaging
of the blue compact galaxy ESO 338-IG04 by \cite{ostlin} who find a
10$^7M_\sun$ SSC of age 6\,Myr.  These objects are sites of intense compact star
formation that can be observed in the optical thanks to the absence of the
thick dust common in nuclei of LIRGs such as NGC\,4418 and Arp\,220.

A young high-mass SSC must have a very large number of Wolf-Rayet (WR) stars whose
existence could in theory be confirmed by optical spectroscopy
\citep{armus1988}.  \cite{armus1989} present optical spectra of NGC\,4418, but
do not detect any WR emission.  The absence of WR signatures could argue
against a young starburst, but could also be explained by the extreme
exctinction (N($\mathbf{H_2)>10^{25}}$\,cm$^{-2}$;\cite{gonzales2012}) towards
the nucleus of NGC\,4418.

\begin{figure}[htbp]
\centering
    \includegraphics[width=0.5\textwidth]{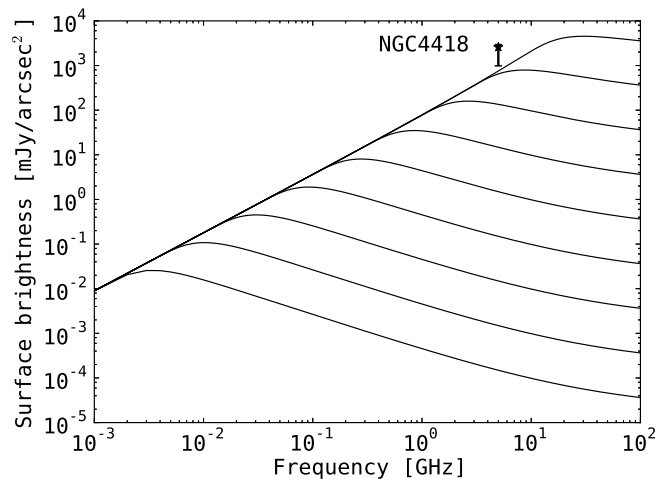}
\caption{
    Radio brightness spectra for star-formation powered radio emission based on
    the mixed synchrotron/thermal model of \cite{condon1991} (see Eq.
    \ref{eqn:condonmodel}) assuming a 10\% free-free emission contribution at
    1.4\,GHz. Successive curves correspond to increasing the frequency at which
    the free-free optical depth is $\tau=1$ which, using Eq.
    \ref{eqn:nuctosfr}, translates to successive curves of star formation rate
    density: 10$^{-3}$, 10$^{-2}$, 10$^{-1}$, 10$^0$, 10$^1$, 10$^2$, 10$^3$,
    10$^4$, and 10$^5$ $[M_{\sun}\mathrm{yr}^{-1}\mathrm{kpc^{-2}]}$. The
    average surface brightness of the resolved features A, B, C, D, E, and H
    detected at 5.0\,GHz in Fig. \ref{fig:wt1.0} is marked with a star.  The
    highest peak surface brightness (A=2970\,mJy/arcsec$^2$) is plotted as the
    upper limit, and the lowest peak surface brightness (E=980\,mJy/arcsec$^2$)
    is plotted as the lower limit.
    \label{fig:condon}
}
\end{figure}

If the nucleus of NGC\,4418 is powered by SSCs with intense star formation, the low
ratio of radio/FIR emission (q=3.075 by \cite{yun}) might be partly explained 
by the thermal absorption of intense star formation as shown in Fig.
\ref{fig:condon}, but also by the star burst being young ($<$5Myear, 
suggested by \cite{sakamoto2013}). Such a young starburst would not yet be
producing the full rate of radio supernovae corresponding to its current star
formation rate as estimated from the infrared luminosity, and hence would 
not fit the FIR-Radio correlation for older starbursts.

The possible attenuation of radio emission at 1.4\,GHz by thermal absorption
was mentioned by \cite{sakamoto2013}. Instead of using the measured 1.4\,GHz
flux density, they extrapolated the total 26.1\,mJy obtained by \cite{baan2006}, using
the VLA at 4.8\,GHz, down to 1.4\,GHz (using $\alpha=-0.8$ typical for synchrotron radiation)
assuming that the 4.8\,GHz emission was not as severely affected by thermal
absorption. They derived a star formation rate (SFR) of
9$M_{\sun}\mathrm{yr}^{-1}$, still too low compared to what is derived from IR, but 
if the starburst is unusually dense we could get significant
thermal absorption even at 5\,GHz. Could thermal absorption explain the low
radio-emission from NGC\,4418?

To model the effect of thermal absorption we use theoretical spectra for a star
forming galaxy using the model of well-mixed thermal/non-thermal emission by
\cite{condon1991} (see Fig. \ref{fig:condon}).
The successive curves are given by the equation
\begin{equation} 
    S_\nu=
    \left(\frac{\nu}{8.4}\right)^2\cdot 10^{-1.3} \cdot T_e \cdot
    \left(1-e^{-\tau}\right)
    \cdot\left[1+10\left(\frac{\nu}{1\mathrm{GHz}}\right)^{(0.1+\alpha)}\right],
    \label{eqn:condonmodel}
\end{equation} 
where $S_\nu$ is the surface brightness in mJy/arcsec$^2$,
$T_e=10^4$K is the thermal electron temperature, $\alpha = -0.8$ is the
typical synchrotron radiation spectral index, and $\nu_c$ is the turn
over frequency where the free-free optical depth $\tau=(\nu/\nu_c)^{-2.1}=1$.

\cite{bell2003} gives the relation between the star formation rate and the
observed 1.4\,GHz radio luminosity as
\begin{equation}
    \mathrm{SFR}[M_{\sun}\mathrm{yr}^{-1}] = 5.52\cdot 10^{-22} \cdot L_{1.4}[\mathrm{W/Hz}].
    \end{equation}
Dividing this equation by 1kpc$^2$ to get surface brightness, we can relate turn-over
frequency $\nu_c$ in Eq. \ref{eqn:condonmodel} at 1.4\,GHz to the
optically thin part of the spectra in Fig. \ref{fig:condon} ($\tau<<1$ in Eq.
\ref{eqn:condonmodel}) as
\begin{equation}
    \mathrm{SFR}[M_{\sun}\mathrm{yr}^{-1}\mathrm{kpc^{-2}]}\approx172\cdot\left(\frac{\nu_c}{\text{GHz}}\right)^{2.1}.
    \label{eqn:nuctosfr}
\end{equation}

The average surface brightness of the resolved features A, B, C, D, E, and H
can be estimated as the fitted integrated flux density divided by the total
area.  Using Eq. 4 in \cite{condon1991} and the sizes from Table
\ref{tab:props}, we get a total area of 7700\,mas$^2$ and a total integrated
flux density of 19.8\,mJy, i.e. an average surface brightness of
2600\,mJy/arcsec$^2$.  Figure \ref{fig:condon} shows this average surface
brightness of features A, B, C, D, E, and H together with theoretical spectra.
The highest peak surface brightness (A=2970\,mJy/arcsec$^2$) is plotted as the
upper limit, and the lowest peak surface brightness (E=980\,mJy/arcsec$^2$) is
plotted as the lower limit.

The average surface brightness of NGC\,4418 falls above the theoretical limit
in Fig. \ref{fig:condon}, but on parsec scales the interstellar medium might be
clumpy and the emission not as well mixed as assumed by this model.
\cite{condon1991} note that "inhomogeneities in real sources will produce
spectral peaks somewhat broader than those plotted". This could explain the
peak brightness of the detected features being up to a factor of three above
the model. 
Another explanation could be if one of the strongest features is due to AGN
activity, e.g. feature A, while the rest of the radio emission is due to star
formation. Unfortunately, the current data do not allow us to clearly separate
any AGN components from the other compact features.

If the features detected in Fig. \ref{fig:wt1.0} are
clusters of star formation described by the model in Fig.
\ref{fig:condon}, the frequency where the free-free optical depth $\tau=1$ would
be close to the upper two curves in Fig. \ref{fig:condon} and we estimate
an average star formation rate per area from Eq.
\ref{eqn:nuctosfr} as $10^{4.5}-10^{5.5}M_{\sun}\text{yr}^{-1}\text{kpc}^{-2}$.
Multiplying by the area of 208\,pc$^2$ for the resolved components A, B, C, D,
E, and H, we estimate the total star formation rate to be 
7-70$M_{\sun}\text{yr}^{-1}$ in the nucleus of NGC\,4418 (possibly even higher
if including the unresolved regions F and G as well).  This is consistent with
previous estimates of 30-100$M_{\sun}\text{yr}^{-1}$ by \cite{sakamoto2013}
(their Fig. 17) using \verb!Starburst 99!  for an instantaneous starburst of
age 1-3Myr, L/M=10$^3L_\sun/ M_\sun$.  Hence the low ratio of radio/FIR
emission might be due to free-free absorption affecting radio emission not only at
1.4\,GHz but also at 5.0\,GHz.
The above free-free absorption model also has the advantage that it naturally explains 
the observed inverted spectrum of the nucleus. 

\section{Summary and conclusions}
\label{sec:conclusions}
Eight compact features have been detected in the nucleus of NGC\,4418 at
5.0\,GHz. The complex morphology and inverted spectrum of the compact features
can be seen as evidence against the hypothesis that an active galactic nucleus
alone is powering the nucleus of NGC\,4418.  This indicates a significant
contribution from star formation.  The compact features can be super star
clusters with intense star formation.  We note, however, that the surface
brightness of the compact features is close to the limit of what can be
produced by well-mixed thermal/non-thermal emission from any surface density of
star-formation. This could be due to an AGN responsible for some of the radio
emission while the rest is due to star formation. Unfortunately, the current
data does not allow us to clearly separate an AGN feature from 
features due to star formation.

New multi frequency radio VLBI observations with the current capabilities of
EVN and eMERLIN are planned in 2014. We hope that these will further constrain
the nature of the compact features in the nucleus of NGC\,4418.

\vspace{1cm}
\begin{acknowledgements}
      E.V., J.C., S.A., and I.M-V. all acknowledge support from the Swedish
      research council.  The European VLBI Network (EVN) is a joint facility of
      European, Chinese, South African and other radio astronomy institutes
      funded by their national research councils.  MERLIN/eMERLIN is a National
      Facility operated by the University of Manchester at Jodrell Bank
      Observatory on behalf of STFC.
      The research leading to these results has received funding from the
      European Commission Seventh Framework Programme (FP/2007-2013) under
      grant agreement No 283393 (RadioNet3).
\end{acknowledgements}

\bibliographystyle{aa}
\bibliography{AA-2013-23303}
\Online

\end{document}